# Comparison of inverse problem linear and non-linear methods for localization source: a combined TMS-EEG study

Ridha jarray[1], Abir Hadriche[1,2], Chokri Ben Amar[1] and Nawel Jmail[2,3].

[1] REGIM Lab, ENIS, Sfax University, Tunisia.
[2] Digital Research Center of Sfax, Tunisia.
[3] MIRACL Lab, Sfax University, Tunisia.

## Abstract:

The Electro-Encephalo-Graphy (EEG) technique consists of estimating the cortical distribution of signals over time of electrical activity and also of locating the zones of primary sensory projection. Moreover, it is able to record respectively the variations of potential and field magnetic waves generated by electrical activity in the brain every millisecond. Concerning, the study of the localization source, the brain localizationactivity requires the solution of a inverse problem. Many different imaging methods are used to solve the inverse problem.The aim of the presentstudy is to provide comparison criteria for choosing the least bad method. Hence, the transcranial magnetic stimulation (TMS) and electroencephalography (EEG) technique are combined for the sake of studying the dynamics of the brain at rest following a disturbance. The study focuses in the comparison of the following methods for EEG following stimulation by TMS: sLORETA (standardized Low Resolution Electromagnetic Tomography), MNE (Minimum Estimate of the standard), dSPM (dynamic Statistical Parametric Mapping) and wMEM (wavelet based on the Maximum Entropy on the Mean)in order to study the impact of TMS towards rest and to study inter and intra zone connectivity.The contribution of the comparison is demonstrated via the stages of the simulations.

## Index Terms:

Brain, EEG, TMS, sLORETA, MNE, dSPM, network connectivity, inverse problem, wMEM

## Introduction:

Although technology has shown significant development until nowadays, studying the brain is still a complex task. Several researchers in different fields of application (signal processors, neurologists, psychologists, biologists) are proposing solutions, indicators and models for studying this organ. Consequently, signal recordings from human brain obtained by an array of electrical sensor using non-invasive electroencephalography (EEG) technique allow the possibility to investigate the complex spatiotemporal activity of the human brain. To study the problem of analyzing the functionality of the human brain and the whole problem of the localizationsource, Transcranial Magnetic Stimulation (TMS) and Electro-Encephalo-Graphy (EEG) are combined to study the relaxation of the brain towards rest following a transient disturbance. More precisely, EEG is one of the best techniques to be used for the study of the human brain and especially for the study of the location and the connectivity of sources in a non-invasive way and with high temporal precision. Therefore, the EEG Neuroimaging technique can estimate the cortical distribution of time varying signals of electrical neuronal activity, and the TMS can target either the median prefrontal cortex (MPFC), which is part of the Default Mode Network(DMN), or the superior parietal lobule (SPL) in the dorsal attention network (Bonnard, M. et al., 2016). Transcranial Magnetic Stimulation (TMS) is a medical technique for transcranial stimulation used in the diagnosis and treatment of certain psychiatric and neurological conditions. Therefore, the Concurrent Transcranial Magnetic Stimulation (TMS) and Electro-Encephalo-Graphy (EEG) are emerging as important tools for assessing cortical properties such as excitation/inhibition, intrinsic oscillatory activity and connectivity (Bergmann et al., 2016; Ilmoniemi et al., 1997; Rogasch and Fitzgerald, 2013; Siebner et al., 2009). The evaluation of various cortical properties such as excitability, oscillations and connectivity is ganged by the simultaneous use of Transcranial Magnetic Simulation with Electro-Encephalo-Graphy (TMS / EEG). Moreover, such combination of methods is technically challenging, resulting in artifacts both during recording and following typical EEG analysis methods, (Nigel C. Rogasch et al., 2017).

Transcranial Magnetic Stimulation (TMS) helps to know how the brain works. In addition, it helps to know howbrain activity controls behavior and how the brain is organized. Researchers also use Transcranial Magnetic Stimulation to find out how brain activity varies in patients with different diseases or conditions. Some of these diseases and conditions include Parkinson's disease, Alzheimer's disease, epilepsy, schizophrenia, autism, traumatic brain injury, chronic pain, stroke etc. Moreover, TMS works by inducing non-invasively and in a focal way electric current in cortical regions. Thus, their activity level is modulated in a variable manner according to the frequency, the number of pulses, the intervals and the duration of stimulation used.

Actually, it is common to detect the same activation zones in PET or fMRI for two different cognitive tasks. The only way to highlight different modes of brain functioning, if they exist, is to study the dynamics of space-time activation of the networks involved. Locating sources in EEG raises many difficulties and is currently the subject of intense research. Such difficulties lie in the forword and inverse problems. Therefore, several works and applications are developed to study the localization source problem (Wang H et al., 2015 ; Jmail et al., 2016, 2017, 2019 ; E. Pirondini et al., 2018; R. D. Pascual-Marqui, 2002; S. Baillet et al., 2001) to pinpoint the source ofanomalies that sense the human brain and to study the network of connectivity between brain regions. In addition, researchers study and evaluate network's connectivity of EEG starting from localization source (forward and inverse problem) to computing connectivity measures. The majority of suchstudies are based on linear methods (MNE, dSPM and sLORETA) and non-linear methods MEM (cMEM, wMEM, rMEM). In the present study, the measures of connectivity of the results of TMS implementations on the brain under EEG signals using two inverse methods which are linear distributed methods: minimum standard estimation MNE (AM Dale et al.,1993); dynamic statistical parametric maps dSPM (RD Pascual-Marqui et al., 2012) and Standardized brain resolution Electromagnetic tomography of the brain, sLORETA(O David et al.,2002) , and non-linear:

maximum entropy on the mean MEM (Cécile Amblard et al., 2004) are studied and compared. These inverse techniques (MNE, dSPM and sLORETA) are considered as distributed methods supposing the same initial assumptions to define active zones with different hypothesis (Jmail et al.,2019). In the same way as the dSPM, at each point the MNE current density map is standardized. While the dSPM calculates the normalization according to the noise covariance, the SLORETA replaces the noise covariance by the data covariance. Thus, the SLORETA manufactures smoother maps in which all zones of the brain potentially activated.

The present work will proceed on the two techniques of linear and nonlinear analysis of localization source or more generally four different methods of localization source: MNE, dSPM, sLORETA, wMEM. The simulation results on a data EEG record following TMS implementations will be compared. The aim of the present study is to compare four methods of inverse problems: MNE, sLORETA, dSPM and wMEM in the definition of network connectivity. a chain pretreatment to assess the rate of connectivity between different regions in the two brain hemispheres have using EEG combined with TMS is used. All simulated models are generated on python and matlab using the Brainstorm toolkit. The present study is composed of three parts: the first part lists the materials used (normal models and the subject to be recorded), the second is devoted to the experimental results and the third part presents the conclusions and the perspectives.

## A/ Materials and Methods:

In order to understand the human brain functions, several research proposed preprocessing schemes to decipher different brain states: rest, cognitive (during reflex), illness. Electro-physiological technique such as Magneto-Encephalo-Graphy (MEG) and Electro-Encephalo-Graphy (EEG) can be applied for the characterization of the functioning of the brain or of these pathologies in a non-invasive manner and with very high temporal precision. Brain activity has been defined in several protocols, by comparing the transient latencies between different experimental conditions and different numbers of sensors. Nowadays, recent progress has made it possible to reconstruct the temporal dynamics in the cortical zones of interest (Mamelak AN et al., 2002). Such reconstructed signals allow their role to investigate the dynamics of network activations (David O et al., 2002). Through connectivity measures, the interactions between the different zones of the brain can be assessed (Horwitz B et al., 2003); (Darvas F et al., 2004)), using measures such as coherence (Gross J, 2001), correlation (Peled A et al., 2001), Dynamic Causal Modeling DCM (Friston KJ et al., 2003), Directed Transfer Function, DTF (Kaminski MJ et al., 1991) but with complex dynamics. Therefore, location source is the solution for determining the zones responsible for excessive discharges. To delimit these regions to be surgically removed, it is necessary to ensure a fine determination of anomaly network, which will serve as a preoperative step. Hence, techniques for evaluating connectivity networks need to be studied and improved.The presentwork attempts to solve the two forward and inverse problems to locate the generators of the stimulus tips while evaluating four problem solving techniques reverse the MNE, the dSPM and the SLORETA and wMEM. Finally, the connectivity obtained by these techniques will be compared.

## 1/ Materials:

All the processing steps have been applied to "Python", "MATLAB" and Mathwork, with the help of Brainstorm toolbox (an open-source collaborative tool for analyzing brain recordings) (F Tadel et al., 2011).The present studied data is an EEG registration of ten right-handed subjects (4 men and 6 women) aged 22 to 30 participated in the experiment. After a brief introduction to the experimental procedures and TMS, all subjects obtained written and fully informed consent. No one has a history of neurological disease or contraindications to TMS.During hardware regulation, TMS compatible equipment used a Magstim 200 stimulator (Magstim Company, Whitland, UK) which generates a monophasic magnetic field of up to 1.7 Tesla, connected to a co-planar figure-eight coil with an outer loop diameter of 9 cm. The coil was held in the desired position by a custom device. The coil could thus be moved or placed in an optimal way, while keeping its stable position throughout the experiment. The stimulation system was connected to a neuro-navigation device (Navigation Brain System 2.3, Nexstim, Helsinki, Finland), using anatomical magnetic resonance imaging (MRI) scans of each subject to precisely guide stimulation. The system calculates an estimate of the electric field induced in the cortex by the TMS pulse in real time (depending on the depth of the site, among others) and displays it on the subject's anatomical MRI

The EEG was recorded continuously with TMS compatible equipment (Brain AMP Direct Current, Brain Products, Gilching, Germany) with a sample rate of 5 kHz, using Brain Vision Recording software. The team used a 62-electrode cap (Fast & Easy Cap) with Ag / AgCl electrode material, mounted on an elastic cap positioned according to the 10–20 method extended to 64 electrodes. In the vicinity of the coil, the cable of each electrode has been tilted so that it is approximately perpendicular to the cable of the coil to limit the artifact induced by TMS on the EEG signal. The characteristics of the electric field induced by each TMS pulse (location, orientation and intensity) were recorded by the neuro-navigation device.Data were segmented into 6 second trials (2 s before, 4 s after TMS). Stimulation artifact was removed in each trial using a custom script cutting the signal from -5 to 10 msec and replacing it with a straight line with added white noise. The signal was then filtered using a 0.5 Hz (order 2) high pass Butterworth filter to correct for any signal drifts.The TMS stimulus is triggered at time t = 0 seconds.

The present study has been authorized by ………………………………………………

## 2/ Methods of localizationsource:

Localizationsource was done by solving forward and reverse problems using the Brainstorm toolbox. To solve the forward problem, a head modelwas used, and then imported the cortex and scalp surfaces to Brainstorm after registration of each MRI subject. A sphere was fixed for each sensor using: nasion, left and right prauricular fiducially markers (L Kossler et al.,2011) after fixing the channel location and then the electrodes with MRI were recorded. For inverse problem, the EEG recordings were imported on Brainstorm after TMS stimulation and then a frequency band filter (50/60 HZ) was used to eliminate well identified contamination from systems oscillating at very stable frequencies, and for the head model a symmetric model (Open MEEG) was chosen, such advanced model uses a symmetric border element method (symmetrical BEM) and also uses three realistic layers (scalp, inner skull, outer skull). The aim of such model is to provide more precise results than spherical models. Then, the other source location techniques MNE, dsPM, sLORETA were performed and MEM for the study data and for the peak groups. A regularization parameter similar to the signal / noise ratio was used. The study's noise covariance matrix is generated as a basic activity before the actual discharges of the events and the noise level at the level of the sensors has estimated.

Distributed methods use a large number of fixed-position "dipole" sources placed on the brain to determine their amplitudes. The gain matrix G connecting the measurements M (t) to the intensity of the sources S (t) is calculated according to the following equation:

$$M(t) = GS(t) + N(t)$$

With N (t) is an additive noise

## 2.1/ MNE:

Minimum Estimate of the standardMNE is a method widely used in solving the inverse problem of source localization (Ana-Sofia Hincapié et al., 2016). It is the basis of a step of regularization of averages, for example the regularization of Tikhonov (Tikhonov A. N et al., 1977).
For the MNE, the measurement matrix should be written as follows:

$$M = R' G^T (G R' G^T + C)^{-1}$$

With G is the gain matrix, C is the noise covariance matrix and R is the source covariance matrix.
The regularization of the MNE results in $R' = R / \lambda^2$

$$M = R G^T (G R G^T + \lambda^2 C)^{-1}$$

The amplitude of the current is described by the term regularization $\lambda$, which has a value inversely proportional to the amplitude of the current.

It should be mentioned, that the MNE is defined by its sensitivity to the surface current. This sensitivity can be attenuated while adjusting the covariance matrix R to value the deep sources. The elements of the covariance matrix R corresponding to the position p are determined by the following equation:

$$f_p = (g_{1p}^T g_{1p} + g_{2p}^T g_{2p} + g_{3p}^T g_{3p})^{-\gamma}$$

$g_{1p}$, $g_{2p}$ and $g_{3p}$ are the 3 columns of the matrix G corresponding to the position p and $\gamma$ is the level of the weighting by the depth.

Many researchers use MNE as a method of inverse problem solving and also for localization source. Among them there are: Olaf et al (Olaf Hauk, 1974) indicate that solving the inverse problem by dipole models requires a priori explicit assumptions about cerebral current sources, which is not the case for solutions based on estimates of minimum norm. Olaf and coworkers assess the spatial accuracy of the MNE minimum norm estimate under realistic noise conditions by estimating its ability to locate sources of evoked responses in the primary somatosensory cortex. They found out that MNE exhibits parametric source modeling results with relatively good localization accuracy and requiring minimal assumptions may be effective in locating poorly understood activity distributions and in tracking changes in activity between zones of the brain as a function of time. Moreover, Hincapié and his collaborators (Ana-Sofia Hincapié, 2016) addressed an interesting question in the field of the localization of MEG sources while using the MNE method around the adjusted threshold of regulation which differs from a cortico-connectivity. It is a cortical to inter-cortical activity.

## 2.2/ dSPM:

The dSPM technique (Dynamic statistical parametric mapping) is another method which allows for compensating for depth bias (Dale AM, 2000) (Liu AK, 2002). The linear inverse operator which is equivalent to a noise-corrected pseudo-inverse operator or Wiener is calculated as follows:

$$P = C_s L^T (L C_s L^T + C_n)^{-1}$$

Where $C_s$ is the source covariance matrix, most assumed the identity matrix. $C_n$ is the normalized covariance matrix.

Then, the normalized operator in noise is calculated, which, in the case of fixed dipole orientations, is given by the following equation:

$$P_{norm} = diag(v)^{-1/2} P$$

With $v = diag(PC_nP^T)$. After the calculation of the inverse operator P, the dSPM performs the noise normalization.

This method is evaluated by several researches like "carrie R and his colleagues". Carrie R and his collaborators (Carrie R. McDonald, 2009) employed a noise-standardized distributed source solution called dSPM in healthy subjects and epilepsy. They verified, in healthy controls, that bilateral visual cortex activity of 80 to 120 ms is restored by dSPM responding to sensory control stimuli and new words. On the contrary, in patients with epilepsy, dSPM can find the timing and spatial extent of language processes. Moreover, Hideaki Shiraishi used the dSPM method to locate epileptiform activity recorded by magnetoencephalography in patients with epilepsy (Hideaki Shiraishi et al.,2005).

## 2.3/ sLORETA:

Another method for depth bias compensation is the SLORETA technique (standardized LOw Resolution brain Electromagnetic TomogrAphy) (Pascual-Marqui RD, 2002). Unlike the dSPM method, the inverse operator is weighted according to the resolution matrix R = PL, linked to the inverse and direct operators "P" and "L".

The pseudo-statistics of power and absolute activation are respectively for the fixed dipolar orientations given by the following equation:

$$\varphi_i = \frac{j_i^2}{R_{ii}} \text{ and } \phi_i = \sqrt{\varphi_i}$$

With P = inverse operator and L = direct operator

The standardized SLORETA inverse operator is given by the following equation:

$$P_{sloreta} = diag\ (r)^{-1/2} P$$

With r = diag (R).

The sLORETA technique was used by Alessia Zarabla in 2017 to locate brain electrical activity and functional connectivity. The aim of the study carried out by Alessia Zarabla and her colleagues was to apply the sLORETA in the evaluation of the possible effects of Ara-C on brain connectivity in patients with AML without CNS involvement (Alessia Zarabla et al., 2017).
Moreover, Michael Wagner and his colleagues used the sloreta method in 2004 to evaluate the performance of the method in the presence of noise and with several active sources simultaneously. They showed that the sLORETA method localizes well, compared to other linear approaches such as MNLS and LORETA. In addition, the sLORETA was used by Martin Brunovskyin 2008 to study cortical sources of EEG in 6239 gray cortical matter voxels (MartinBrunovsky et al., 2008).

## 2.4/ MEM:

The MEM is a non-linear method of localizationsource. This technique is based on the estimation of sources J as a multivariate random variable of dimension p, with a probability distribution dp (j). Moreover, the MEM principle aims at estimating the distribution $d\hat{p}(j)$ which provides "a maximum uncertainty on the missing information carried by the data" (Jaynes ET, 1957), compared to a reference model assumed on J (Amblard C et al., 2004). Regularization in this context is introduced by the drafting of the solution in the form of dp (j) = f (j) dv (j), where the reference distribution dv expresses some hypotheses on J and f (j) is a density of v.

To find that, the data on average is explained as follows:

$$M = \int Gj\, f(j)\, dv(j)$$

The solution MEM $d\hat{p} = \hat{f} dv$ is the one with maximum v-entropy. An interesting property of the MEM approach lies in itsflexibility inherent in introducing constraintsdefinition of the reference distribution dv.

In the present study, dv was defined using the P (s) parcelization model by assuming that K represents cortical plots showing a homogeneous state of activation describe brain activity. The global dv was defined as follows:

$$dv(j) = dv_1(j_1)dv_2(j_2)\ldots dv_k(j_k)\ldots dv_K(j_K)$$

MEM can also be applied in the representation of wavelets, where the time index is modified for the time-frequency indices. We talk here of wMEM (MEM in the Wavelet Domain): MEM in the wavelet domain. Denote as p (w) the joint probability of the wave coefficients of all sources, at a particular direct time (k) and scale (j). In this part, the indices k and j have been removed for clarity. The MEM estimate deduces the expectation Ep [w] assuming a reference probability μ (w) from which the entropy deviation is minimized under the constraint of fit data.

The wMEM technique was used by Md. Shakhawat Hossain in 2017 on BCI Competition EEG data for classifying two motor imagery (MI) tasks using optimal electroencephalography (EEG) sources. Md. Shakhawat and his colleagues reported that the best classification accuracy achieved is 98% using only 11 selected channels, which is close to 100% achieved using the 118 available channels. Such result summarizes how the optimal EEG channels can be used to develop a BCI system without significantly compromising performance. In addition, Simanto Saha and colleagues used the wMEM method in 2019 on EEG data for the purpose of identifying inter-subject associative electromagnetic sources, estimated from the wavelet-based maximum entropy of the mean (wMEM) of the l 'Single test EEG, and project them into a three-dimensional (3D) head model.( Simanto Saha et al.,2019).

JM Lina used the wMEM technique in 2014 to describe, evaluate and illustrate with clinical data a new method of localizing the sources of the oscillatory cortical activity recorded by MEG and to address the question of the localization of events of a single test, which are typically associated with low signal-to-noise ratios (SNRs) (J.M.Lina et al 2014).

## 3/ Criteria of comparison:

This part presents the suggested methodology for comparing the other distributed MNE, dSPM, sLORETA and wMEM localization source methods for studying the impact of TMS technique on EEG data.

## 3.1/Cross-correlation and Connectivity:

At the source activation point, several activation films were recorded by the different methods (MNE, dSPM, sLORETA and wMEM). Then, the active regions were visually determined, and the regions of interest on the cortex named "Scouts" were defined. The active zones for the oscillations in the two right and left hemispheres were selected by setting five scouts to each. The time series for each region of interest was restored, using the singular value decomposition following the projection on the regional dipoles in each of the methods (figure 1).

Figure 1 explains the temporal declines at the source level for the patient oscillations of the different localization methods (MNE, dSPM, sLORETA, wMEM).

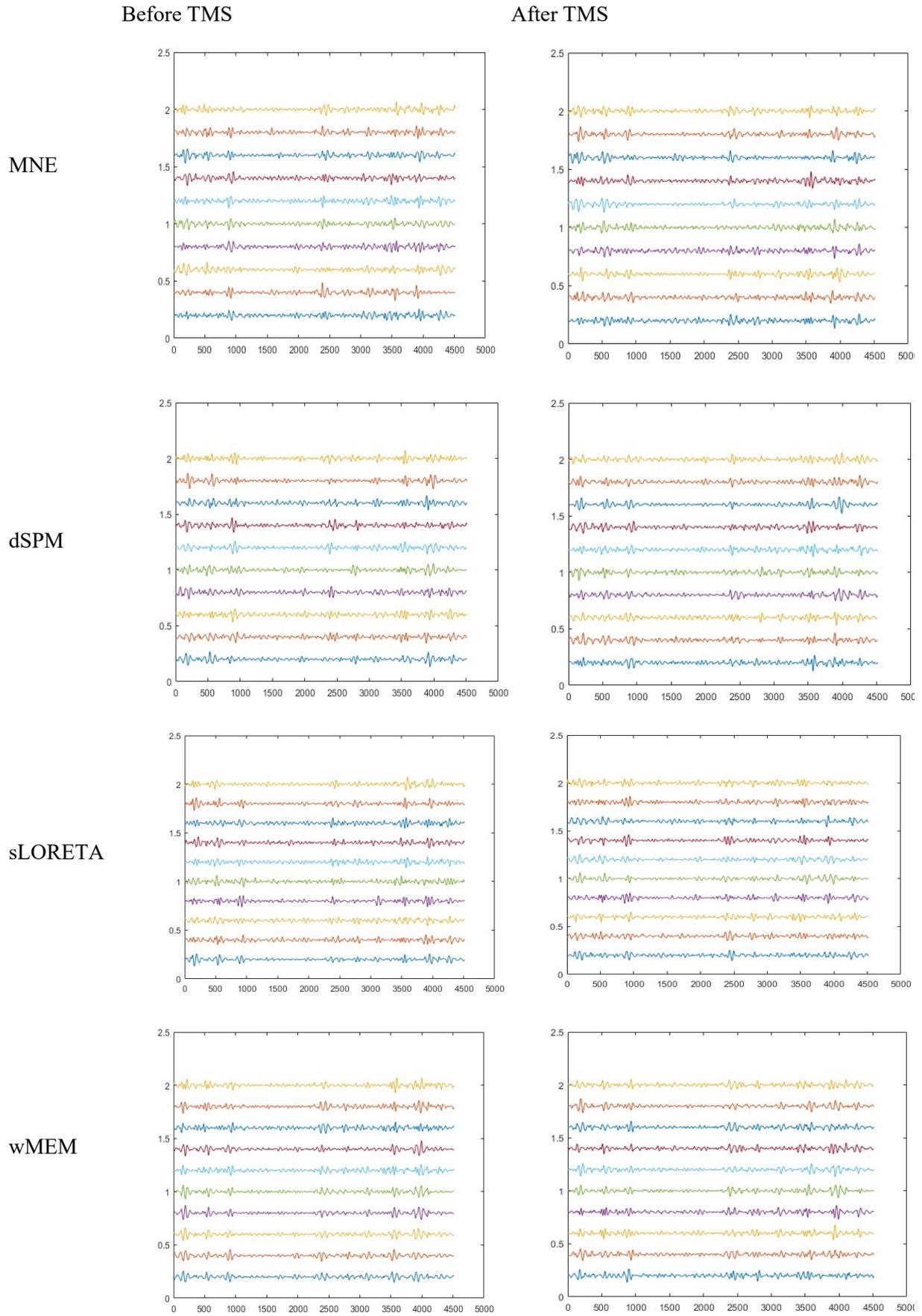

*Figure1: Scout time series of both MNE, dSPM, sLORETA and wMEM methods before and after using TMS*

MNE

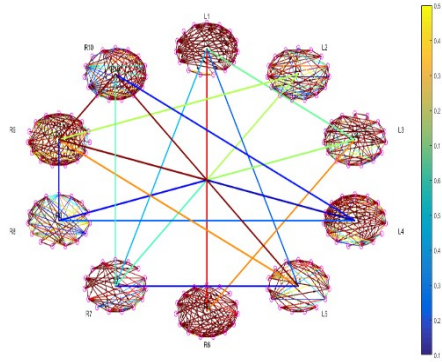 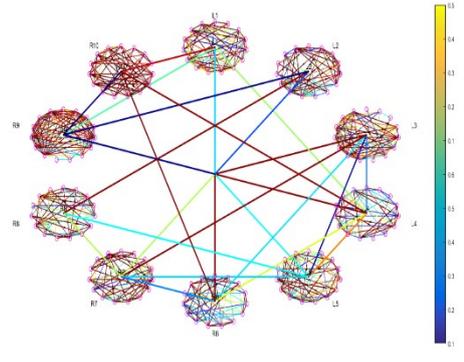

dSPM

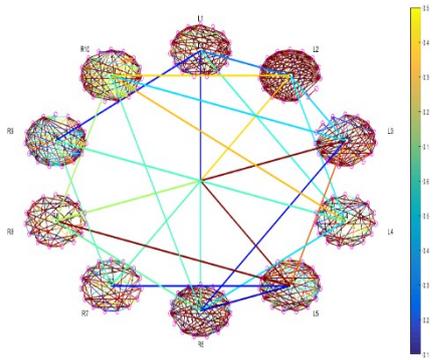 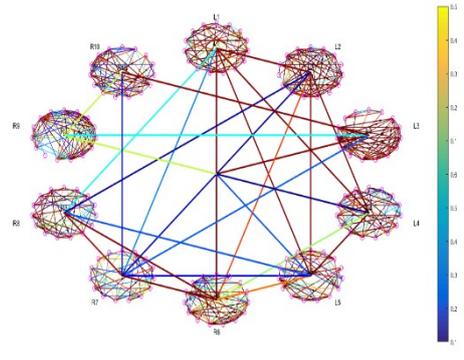

sLORETA

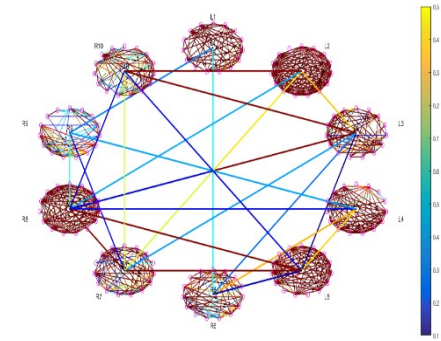 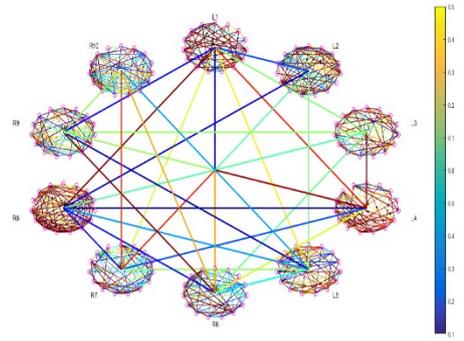

wMEM

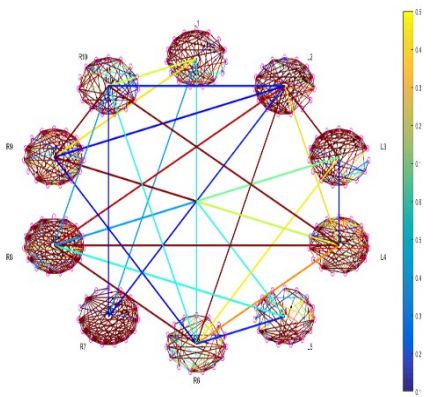 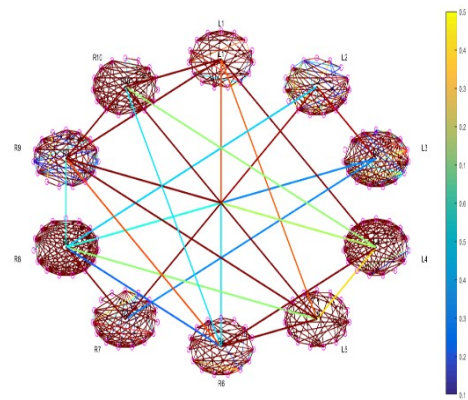

Before TMS            After TMS

MNE

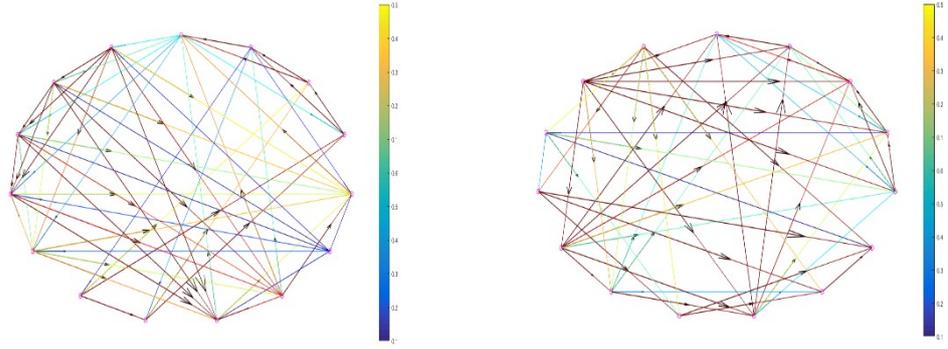

dSPM

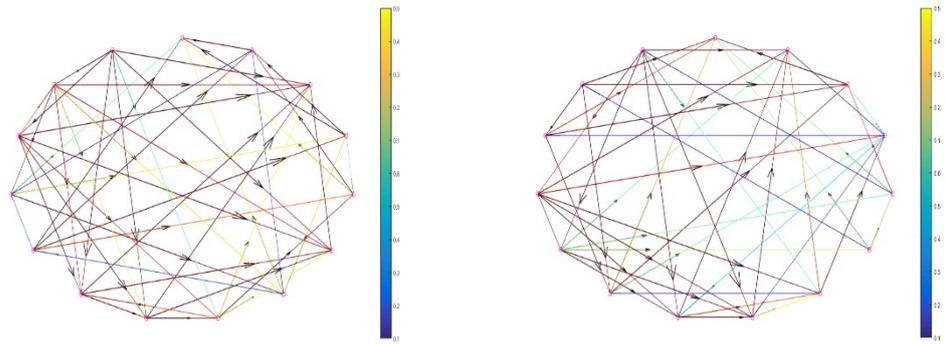

sLORETA

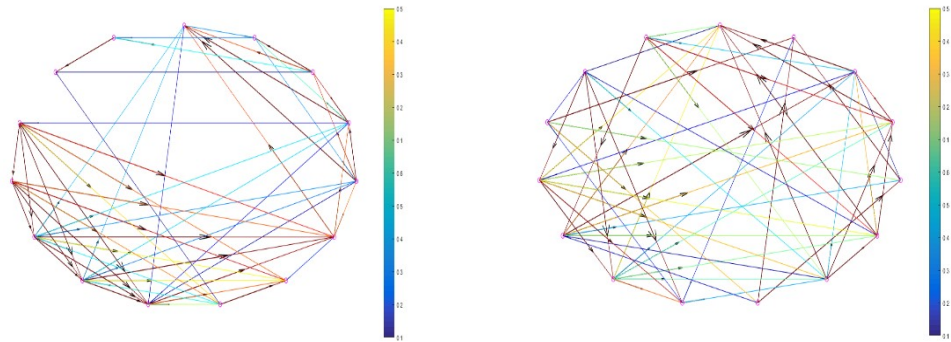

wMEM

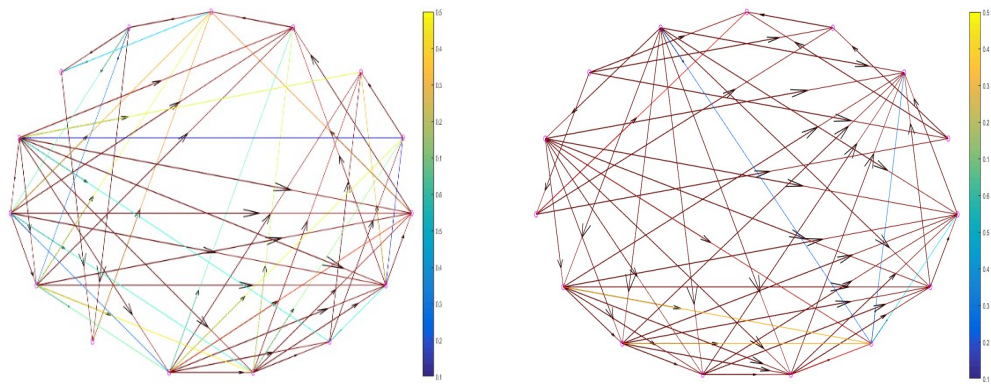

*Figure2: The cross-correlation between these time paths for each signal pair of the MNE, dSPM, sLORETA and wMEM methods.*

For each pair of signals, the cross correlation between these temporal pathswas calculated in order to be able to assess the relationship between these active zones using the study four distributed inverse techniques (Figure 2).

The level of correlation between different channels is indicated by the line colors. The red line corresponds to a strong correlation and the blue line corresponds to a weak correlation. It is observed from the study results that the MNE showed a weak correlation between the active cortical regions since the number of blue links is greater compared to the number of red links. But the dSPM method shows a strong connection compared to the MNE method and that the number of red and blue links is almost the same, for this reason we say that dSPM is said to have an average correlation.

The evaluation of the coupling rate between the selected active regions in the two hemispheres (right and left) using the sLORETA and the wMEM methods shows the presence of many of the connections between the different circles (all regions are connected to each other). Also, the number of connectivity links increases after using TMS. This is why it is said that the sLORETA and the wMEM methods have a strong correlation and on all that the sLORETA method produces smoother maps in which all zones of the brain potentially activated and due to the low spatial resolution, all regions are connected, regardless of the depth of the sources.

## 3.2/Connectivity inter and intra zones:

To evaluate the inter and intra zone connectivity, the global connectivity indicators (index of kansky α, β, γ ) (K.J.kansky,1989) were used to calculate and study the connectivity before and after the use of TMS for the different networks traced by the methods sLORETA, MNE, dSPM and wMEM:

- $\beta = \frac{e}{v}$, very summary indicator comparing the number of edges to the number of vertices,e and v represent respectively the number of edges and the number of vertices
- $\gamma = \frac{2e}{v(v-1)}$, provides information on the density of edges in the network
- $\alpha = \frac{2(e-v+p)}{(v-1)(v-2)}$, is the connectivity index, p representnumber of subgraphs

*Table 1: global indicators of inter-zones connectivity before and after TMS for the different methods of source location.*

|  | Before TMS | | | | After TMS | | | |
|---|---|---|---|---|---|---|---|---|
|  | MNE | dSPM | sLORETA | wMEM | MNE | dSPM | sLORETA | wMEM |
| Edges (e) | 21 | 26 | 25 | 27 | 25 | 28 | 33 | 29 |
| Vertices (v) | 10 | 10 | 10 | 10 | 10 | 10 | 10 | 10 |
| Bêta (β) | 2.1 | 2.6 | 2.5 | 2.4 | 2.5 | 2.8 | 3.3 | 2.8 |
| Gamma (γ) | 0.46 | 0.57 | 0.55 | 0.6 | 0.55 | 0.62 | 0.73 | 0.64 |
| Alpha (α) | 0.33 | 0.47 | 0.44 | 0.5 | 0.44 | 0.53 | 0.64 | 0.56 |

*Figure3: intra-zone connectivity of the MNE, dSPM, sLORETA and wMEM methods.the links of different colors show the connectivity between the 15 vertices for a scout, the arrows colored in black show the sense of directionality.*

To assess the intra-zone connectivity, the number of vertices was increased to 15 vertices for each scout, then the connectivity between the vertices was plotted. (see figure 3).

*Table 2: global indicators of intra-zones connectivity before and after TMS for the different methods of source location.*

|  | Before TMS | | | | After TMS | | | |
| --- | --- | --- | --- | --- | --- | --- | --- | --- |
|  | MNE | dSPM | sLORETA | wMEM | MNE | dSPM | sLORETA | wMEM |
| Edges (e) | 58 | 49 | 55 | 55 | 51 | 47 | 53 | 53 |
| Vertices (v) | 15 | 15 | 15 | 15 | 15 | 15 | 15 | 15 |
| Bêta (β) | 3.87 | 3.27 | 3.66 | 3.66 | 3.4 | 3.13 | 3.53 | 3.53 |
| Gamma (γ) | 0.55 | 0.47 | 0.52 | 0.52 | 0.49 | 0.45 | 0.5 | 0.5 |
| Alpha (α) | 0.48 | 0.38 | 0.45 | 0.45 | 0.41 | 0.36 | 0.43 | 0.43 |

As it is shown in Table 1, the value of the alpha connectivity index increases at the inter-zone connectivity level for all the MNE, dSPM, sLORETA and the wMEM methods after TMS compared to the value of alpha before TMS. But at the level of intra-zone connectivity, the value of the alpha index is reduced after TMS compared to the result of alpha before TMS (see table 2) for the linear (MNE, dSPM, sLORETA) and non-linear (wMEM) methods. At this point it can be seen that the TMS technique has a great influence on the strength of connectivity between the two hemispheres of the human brain.

### 3.3/method of detecting active regions:

The K-Means algorithm was used for the segmentation and the detecting step of the active zones before and after the use of TMS according to different source location methods, either linear methods (dSPM, MNE, sLORETA), or the non-linear method (wMEM).

The implementation of the study result, which is used to detect active zones, goes through the following steps:

-Fixing the source location result capture time (-120ms before using TMS and 33ms after TMS).

-Segmentation of the images according to the different colorations of the active zones by the K-Means algorithm.

The results of segmentations by the K-Means algorithm are illustrated as follows: (see figure 4)

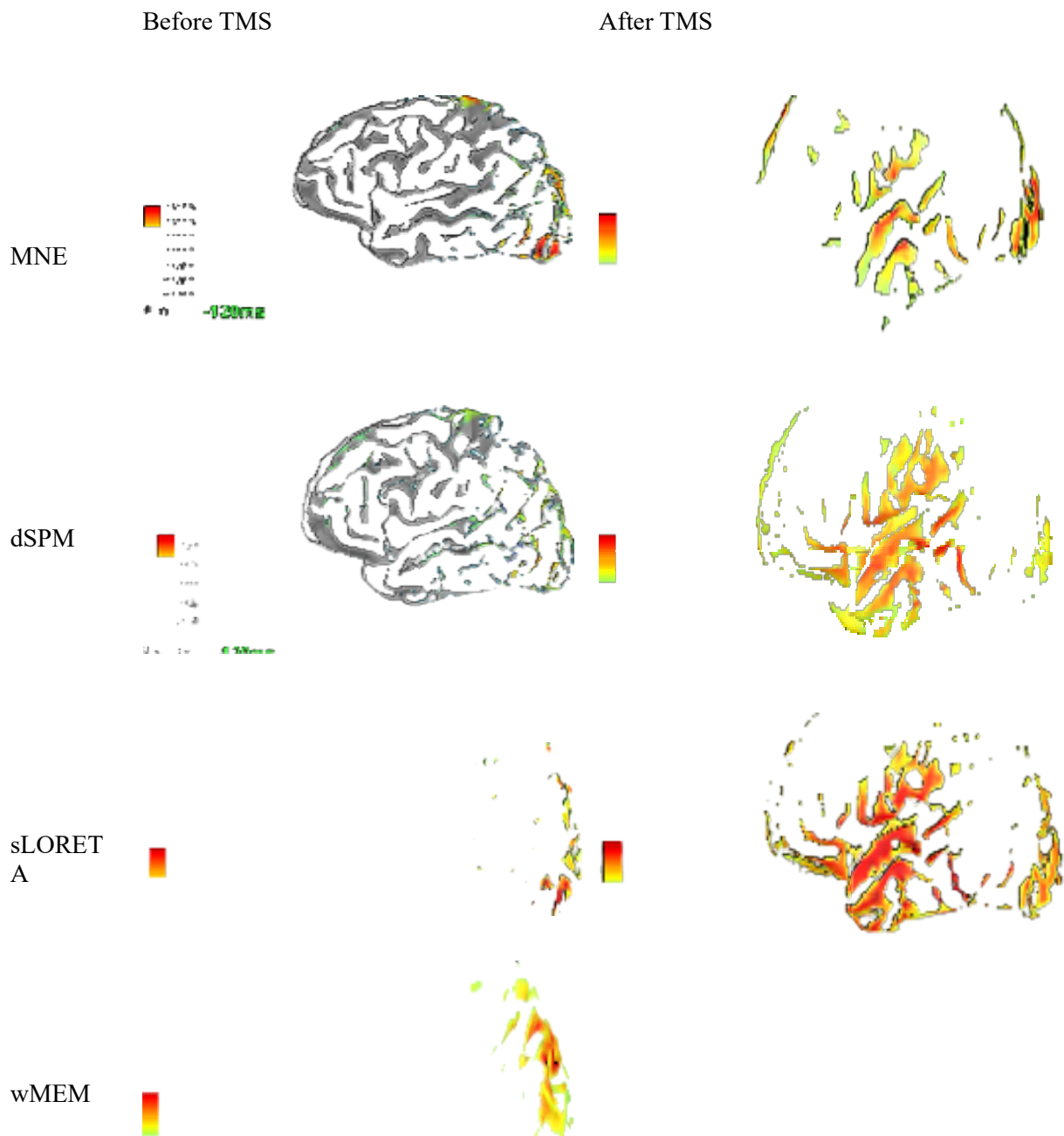

*Figure4: Detection of active zones before and after TMS pulse using MNE, dSPM, sLORETA and wMEM methods.*

It is visually noted that in all the results, when using a linear method of source location (MN, dSPM, sLORETA) or non-linear (wMEM) that the number of activated zones increases after the use of TMS compared to the result before TMS. However, the results show that in all the linear or non-linear source localization methods the number of activated regions increases after TMS and also that the active regions are found at the level of the occipital lobe of the brain, which represents the alpha wave.Such results show that the disturbance by TMS increases the alpha rhythm(Bonnard et al., 2016)

In addition, the conclusion in a visual way is not taken into account and always remains a mere conclusion, which differs from person to person. This is why the detection rate of active zones in the different source location methods, either linear or non-linear was calculated.

Based on figure 4, the different methods only detect the active zones in the occipital lobe, but after TMS stimulation the number of active zones is increased and the detection is in almost all lobes (occipital, frontal, parietal) which means that TMS has been applied near the medial line.The dominance of dark red and yellow-orange color identifies the different active regions in the brain cortex.

The apparent disparity in the number of active regions in the different location methods shows the extent to which the method is capable of identifying actives zones.The dark red and yellow-orange color zones identify the active zones detected by the methods distributed in the human brain of source localization.

To calculate the detection rate of active zones the method of extracting each color in number of pixels and in percentage was used.

The calculation result is summarized in the table 3:

*Table 3: active regions detection calculation rate*

| Method of localization | Calculation rate before TMS | Calculation rate after TMS |
|---|---|---|
| MNE | 0.8% | 1.28% |
| dSPM | 0.58% | 3.7% |
| sLORETA | 1.71% | 5.15% |
| wMEM | 1.54% | 2.78% |

According to the study results, first, the detection rate of active zones increases after the use of TMS compared to the rate before the use of TMS in the different methods used.

Second, the dSPM method performs better than MNE and the wMEM, but not as well as sLORETA for the detection of active areas following TMS stimulation. Third, the sLORETA is superior to wMEM, dSPM and MNE in terms of detecting active zones between a reference region in the brain cortex (after an TMS stimulation procedure).

Fourth, the sLORETA method has a higher detection rate, more active scouts (high connectivity) than other methods; therefore, sLORETA is the most capable method of locating active zones.

To get a fair comparison, the source localization results for each method after using TMS at the start and middle and end were tested (t = 0.12s, t = 0.22s, t = 0.33s, t = 0.5s). (see figure 5)

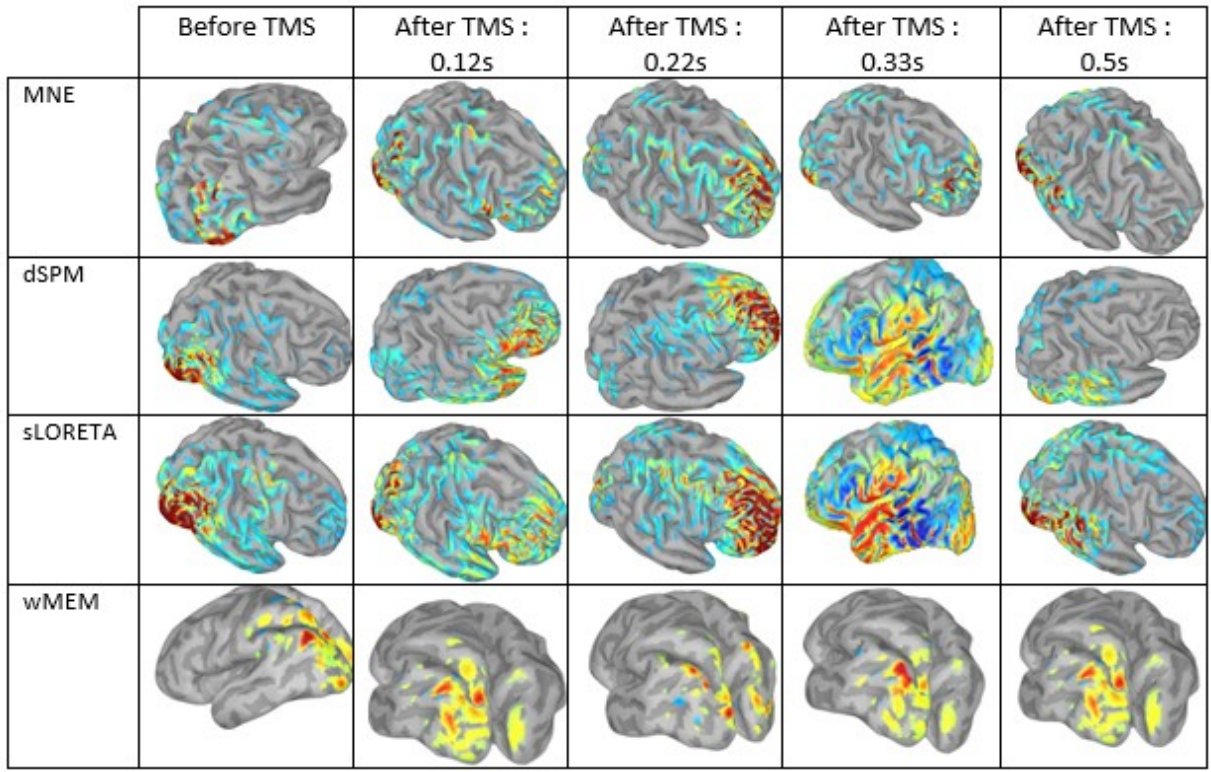

*Figure 5 : Activation of zones at different time T.*

The stimulation results show that before the use of TMS in all distributed methods the active areas are at the level of the occipital lobe but after the TMS there is a decrease in the number of active areas at the onset of TMS.Then there is an increase.And from 0.4s the number of active areas decreases at the end of TMS in all four methods and returns earlier to the initial state (before TMS: activation of areas are located just in the occipital lobe).

It is also noticed that the activation of the regions is higher in the frontal lobe and in the occipital lobe which means that stimulation by TMS creates a strong coupling between the occipital and frontal lobe.

## B/ Results and discussion:

In this part the results of the simulations obtained in each of the source location methods used either the linear or non-linear methods will be analyzed:

- **MNE method:**

Figure 6 illustrates the left and right part of the brain obtained by the MNE method before TMS.

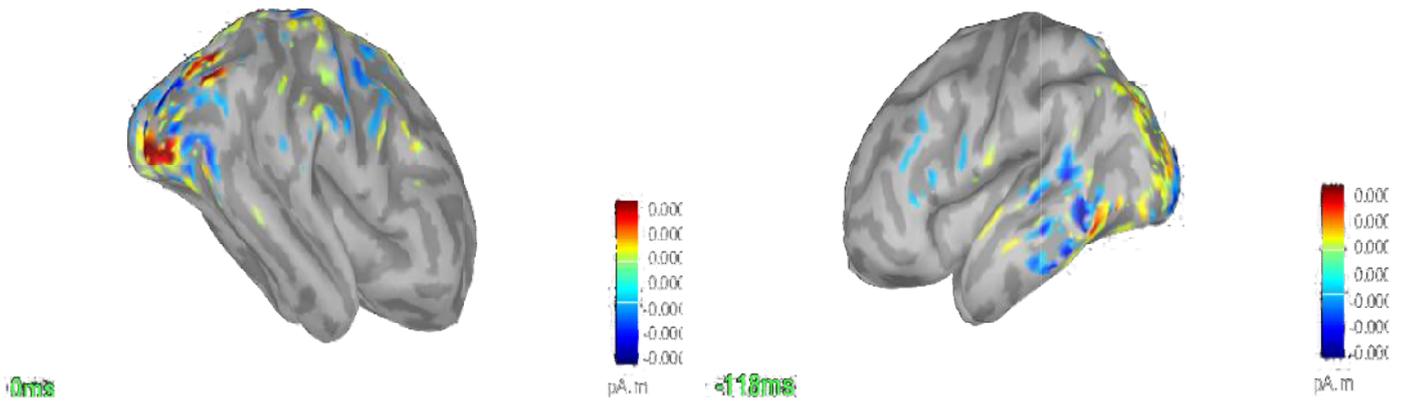

Figure 6: the left and right part of the brain obtained by the MNE method before TMS

In figure 7, the right part of the brain obtained by the MNE method after TMS is illustrated.

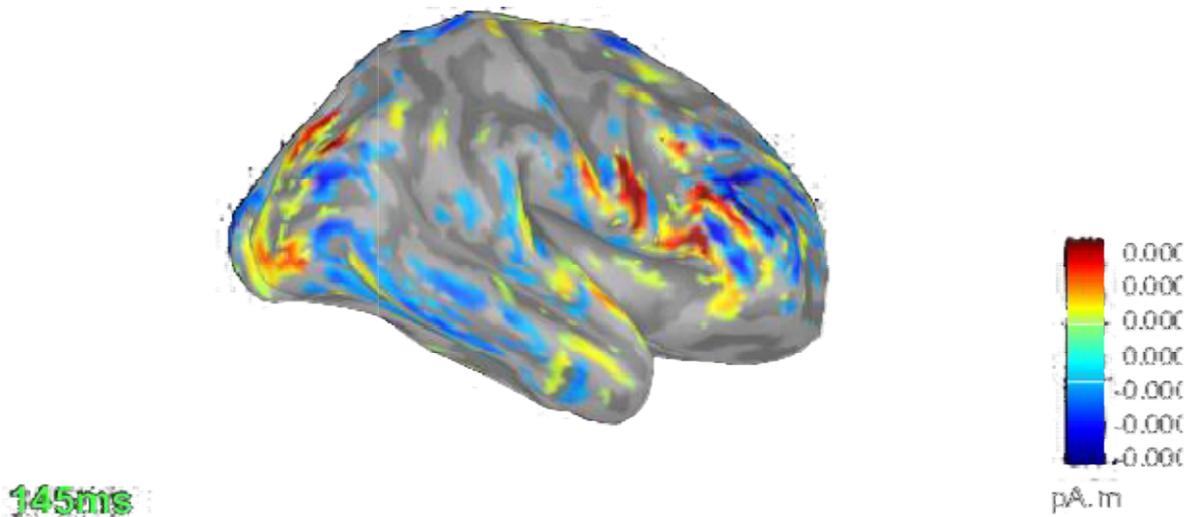

*Figure 7: the right part of the brain obtained by the MNE method after TMS*

For the MNE method, an activation film was obtained. Visually active zones have been defined.

In Figures 8 and 9, the active regions of the study EEG data before and after TMS using the MNE method, with a threshold "p = 25" are represented.

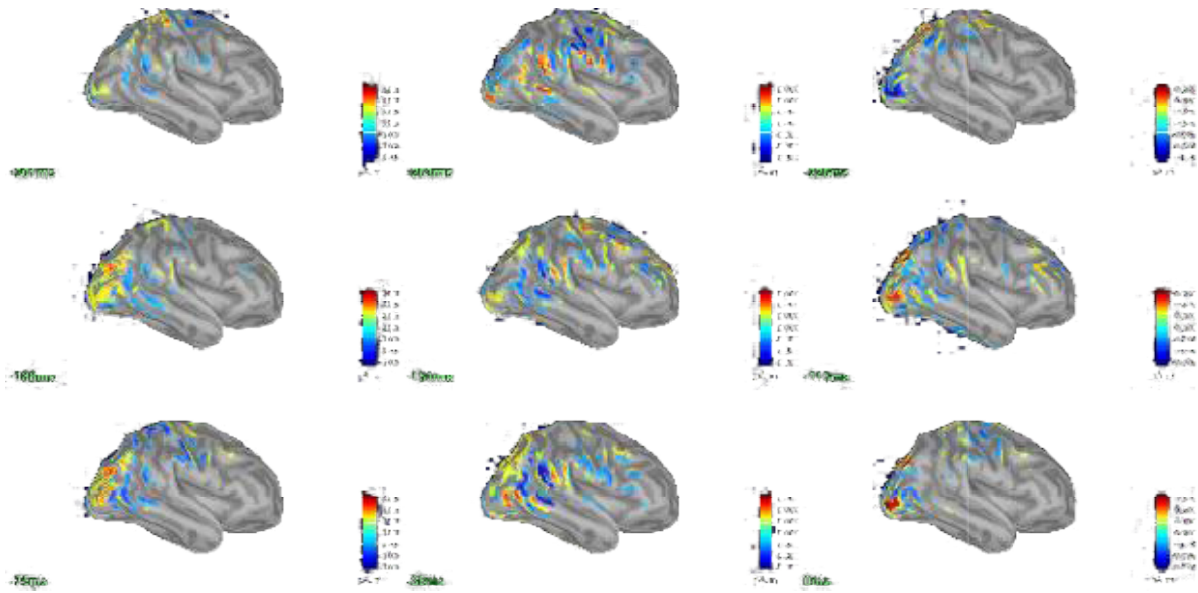

*Figure 8:300 ms source activation film for the right hemisphere point group before TMS*

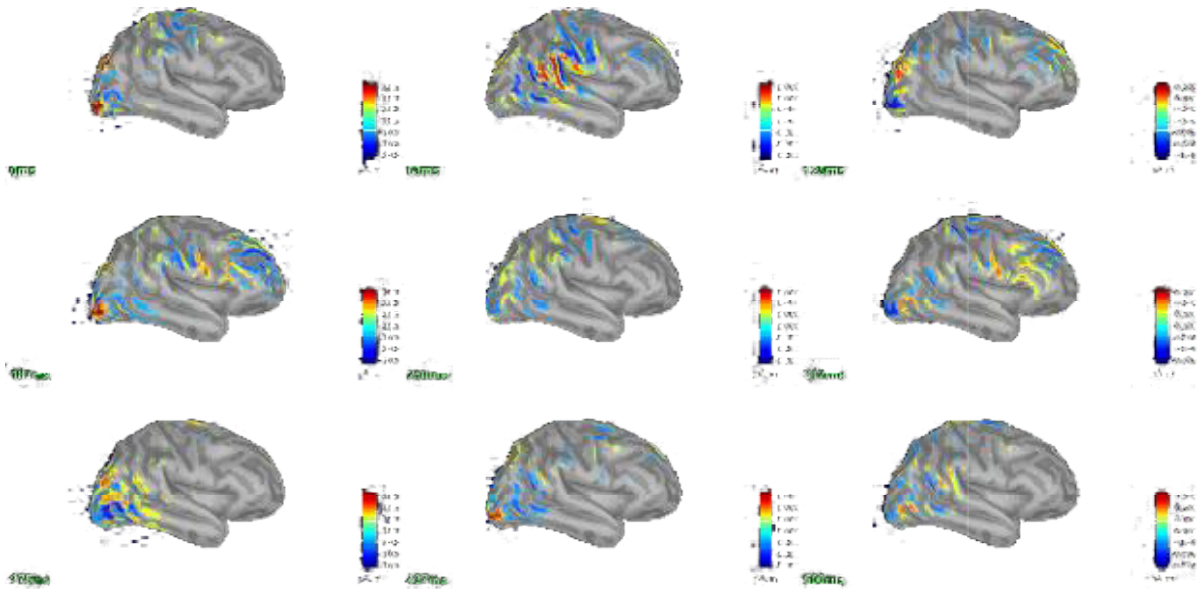

*Figure9:500 ms source activation film for the right hemisphere point group after TMS*

The choice of MNE as a technique for solving the inverse problem was justified by the fact that this technique does not a priori require an exact number of sources. However, it requires regularization, which can have an effect on the estimation of time series, in the form of cross talk between sources. It can then be profitable to use a constraint of parsimony on the sources (David O, 2002). The MNE solution using sources with constrained orientations orthogonally to the cortex, with a noise signal equal to 3 which is used in the level of regularization, a weighting by the depth of a factor of around 0, 5 (these characteristics have been validated by an expert neurologist) was calculated. Then, the inverse calculation from the gain matrix was applied in order to determine for each regional source, three time courses per region. Finally, decomposition into singular values was carried out and the component which has the most energy was selected.

It is noteworthy that in the activation film after the TMS the presence of several active regions and these regions are all connected to each other with respect to the activation film before the TMS.

- **dSPM method:**

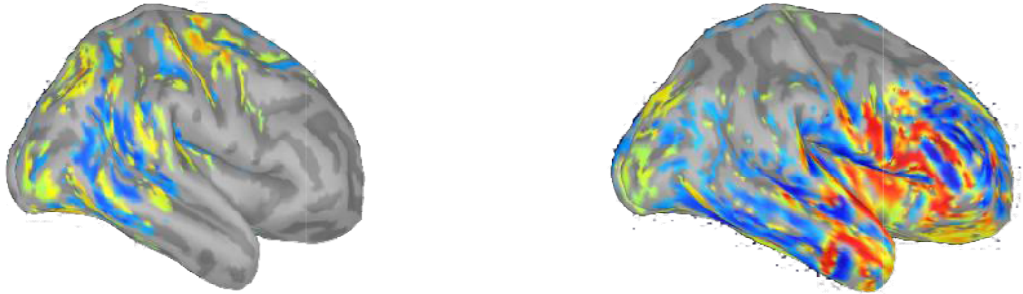

Figure 10: Active zones by dSPM, the right hemisphere. Left: before TMS, Right: after TMS.

Figure 11 and 12 is a film depicting activation of the cortical regions by "dSPM" as a function of time before and after TMS.

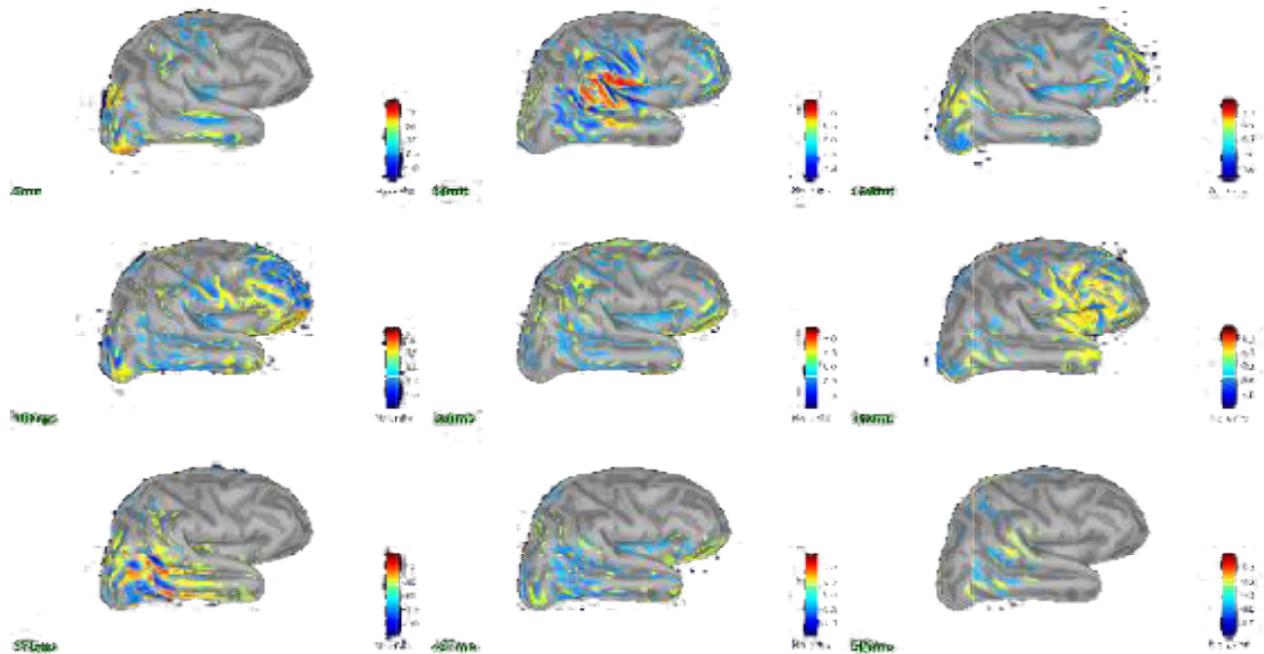

Figure 11: 300 ms source activation film for the right hemisphere point group before TMS.

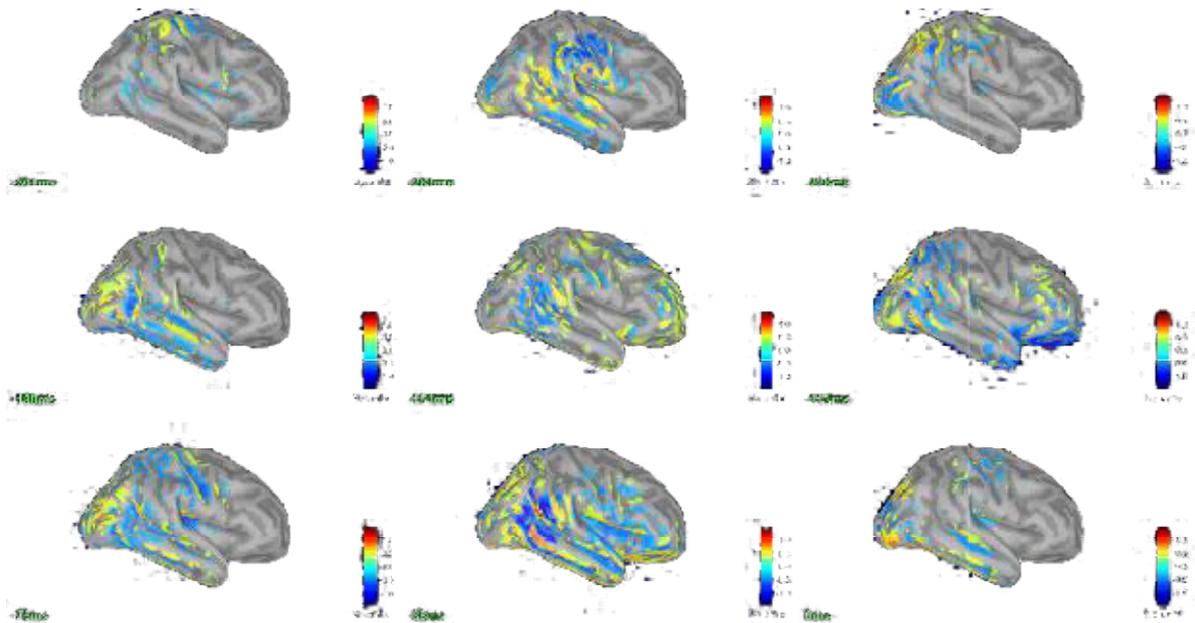

*Figure12: 500 ms source activation film for the right hemisphere point group after TMS*

Although there are many active zones since the simulation result (activation film) of the dSPM method after the TMS compared to the result before the TMS but it can be observed from the activation film that the "dSPM" method shows a weak connection because not all zones are connected to each other.

- **The sLORETA method:**
  Low resolution normalized electromagnetic tomography of the brain is a localization method using the covariance of the data in the case of normalization calculations. Figure 13 represents the location of the source of active zones by the sLORETA method before and after TMS i.e. the active zones.

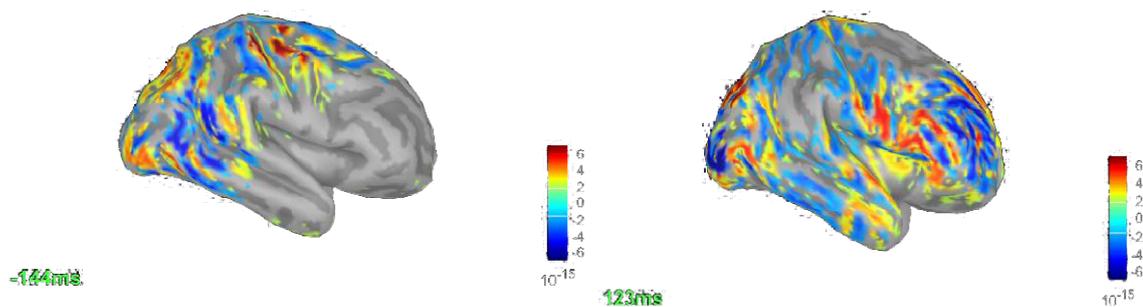

*Figure13: Active zones by SLORETA, Right: after TMS, Left: before TMS*

An activation film by sLORETA was created. The results of localization of the sources of the active zones are illustrated in figures 14 and 15.

Figures 14 and 15 represent a film, which describes the activation of the cortical regions by "sLORETA" as a function of time respectively before and after TMS.

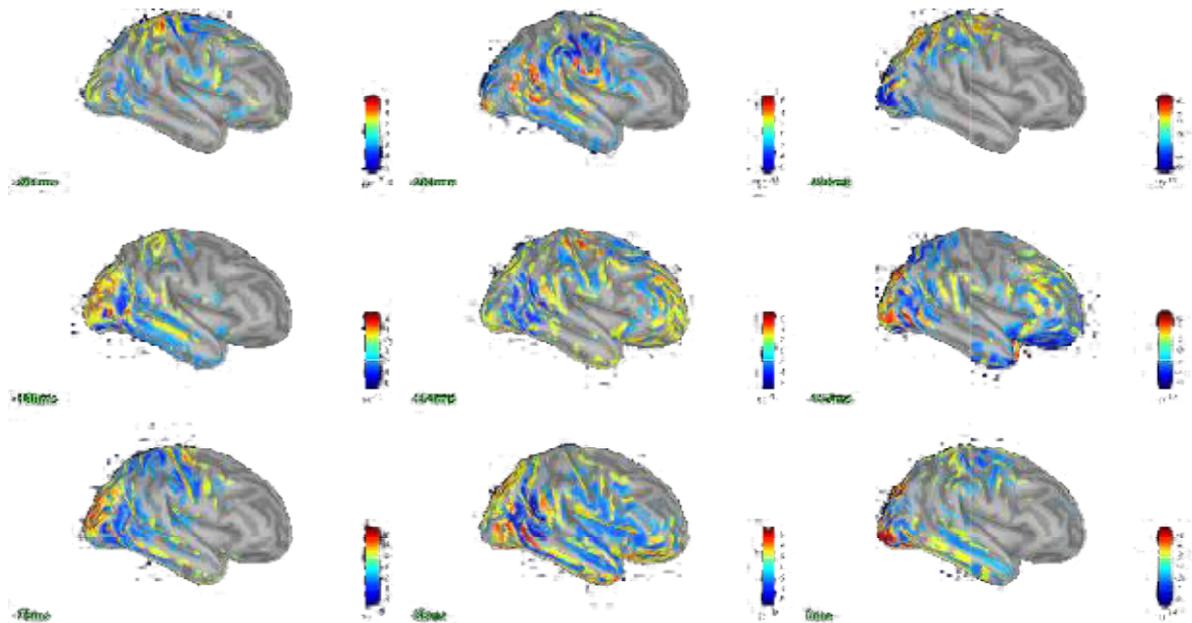

*Figure14: 300 ms source activation film for the right hemisphere point group before TMS*

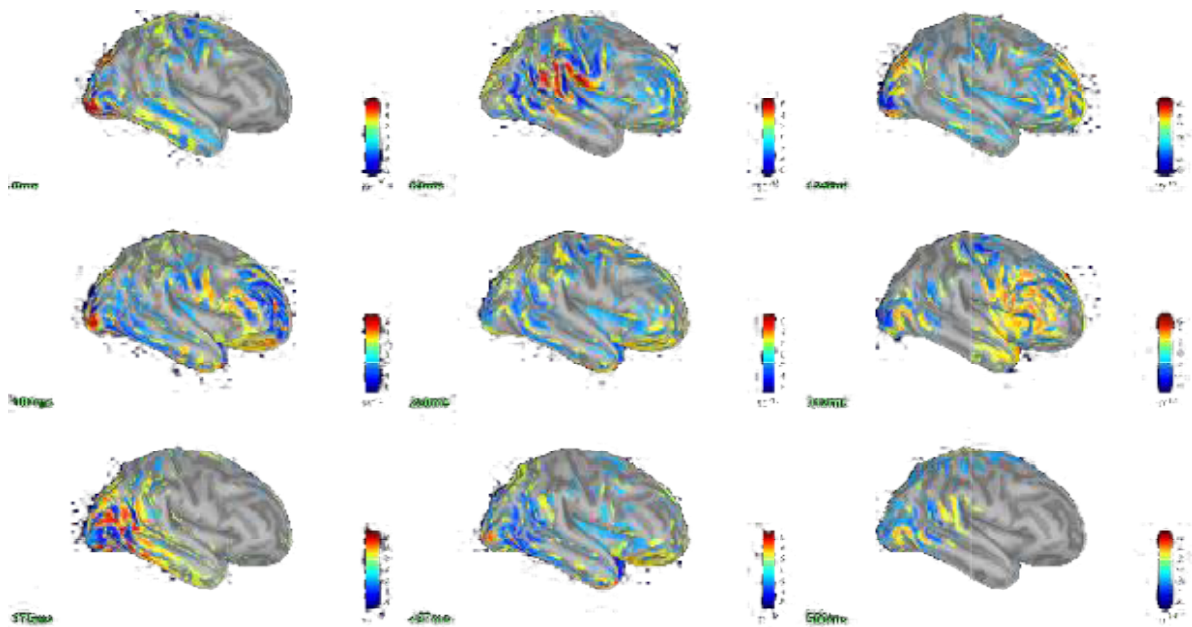

*Figure15: 500 ms source activation film for the right hemisphere point group after TMS*

Using the "sLORETA" method, the presence of a strong connection between the selected active regions in both hemispheres (right and left) is shown. Consequently, thesLORETA method makes smoother maps in which all zones of the brain potentially activated. Due to the low spatial resolution, all regions are connected, regardless of the depth of the sources.

- **The non-linear MEM method:**

Maximum Entropy on the Mean is a nonlinear method of source localization based on a probabilistic (Bayesian) approach where the intensities of the current source are estimated from the informational content in the data (notion of maximum entropy). There are 3 types of MEM: wMEM (time-scale representation), cMEM (time series representation) and rMEM

(wavelet representation). The present study focused on the wMEM type. Then, the wMEM wavelet based on MEM is dedicated to perform source localization of oscillatory patterns in the time-frequency domain.

Thefollowing table displays the simulation result of the study EEG data before and after TMS implementations using the wMEM method. All the results displayed in the table are obtained after the regulation of the following parameters (frequency, wave type and threshold). Therefore, the threshold value was set "p=25" and the frequency and the wave type were changed.

*Table 4: Activation of zones before and after using TMS for the different waves.*

| frequency | Wave type | Result before TMS | Result after TMS |
|---|---|---|---|
| 7.5 : 500 Hz | All wave | 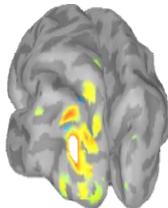 | 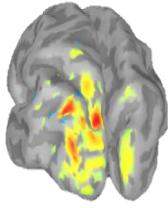 |
| 31 : 62 Hz | Gamma (γ) | 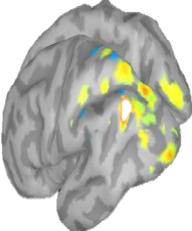 | 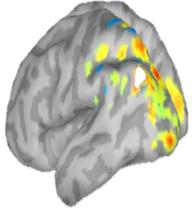 |
| 15.5 : 31 Hz | Beta (β) | 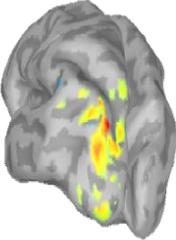 | 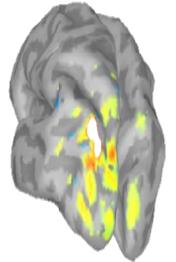 |
| 7.5 : 15 Hz | Alpha(α) | 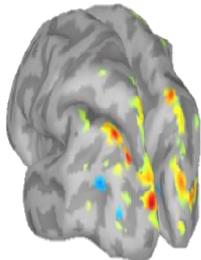 | 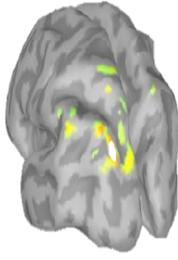 |

It was found that the active zone in the brain is always located at the parietal lobe and at the occipital lobe. That is to say that the wMEM method always locates the alpha wave whatever the frequency used. In addition, the wMEM then locates the sources associated with each wavelet coefficient of the time scale plan and thus gives an estimate of the wavelet coefficients for each source and calculates a scale-specific covariance matrix and a scale-dependent shrinkage.

Based on the obtained results, it was also found that that the wMEM method is performed regardless of the frequency scale used (there are active zones in all frequency scales).

Therefore, the oscillatory activities of the brain measured using EEG signal following TMS impulsionshave been located and based on the assumption that this activity results from processes occurring at different time scales (i.e. bands frequencies) located in certain large cortical regions.

The wMEM is based on wavelets dedicated to the localization of source of oscillatory patterns in the time-frequency domain and on a probabilistic (Bayesian) approach and more generally on the notion of entropy.

Therefore, the estimation results of entropy and the different frequency value are summarized in the following table:

*Table 5: The results of the entropy estimation and the different frequency values for the wMEM method.*

| frequency | Wave type | Entropy | Multi-Resolution power | Wavelet coefficients |
|---|---|---|---|---|
| 7.5 : 500 Hz | All wave | 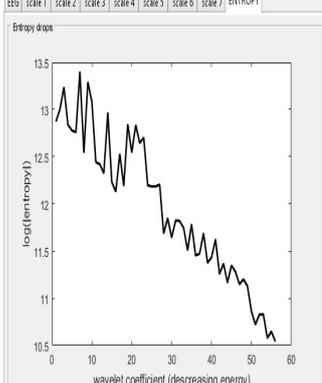 | 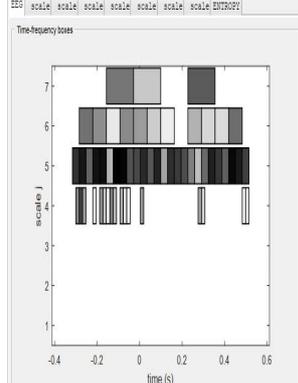 | ---------- |

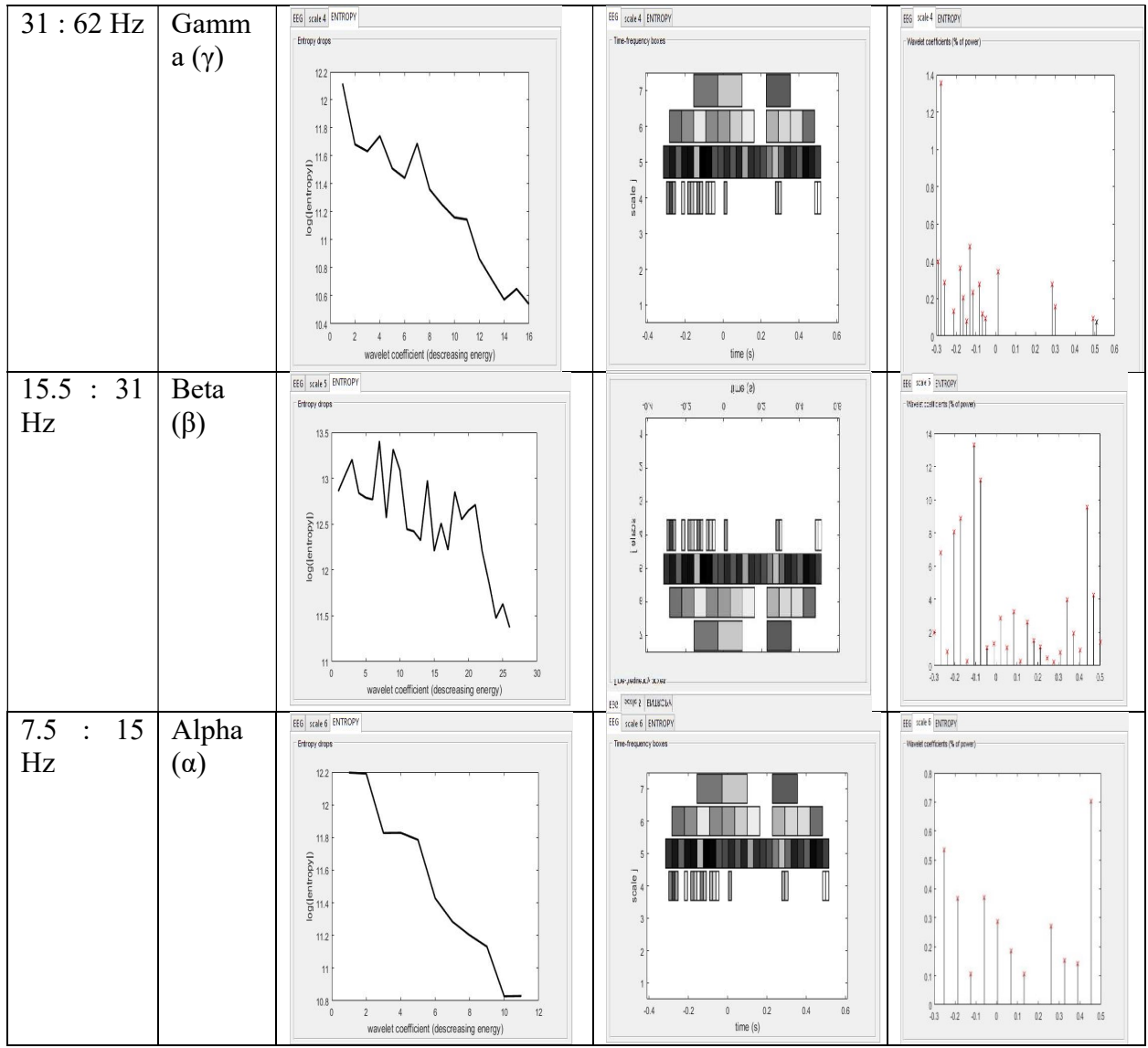

| 31 : 62 Hz | Gamma (γ) | | | |
| 15.5 : 31 Hz | Beta (β) | | | |
| 7.5 : 15 Hz | Alpha (α) | | | |

The figures in the "Multi-Resolution power" boxes show the overall multi-resolution power (average power in wavelets on all EEG sensors). Each box corresponds to a wavelet on a certain scale (index j) and located in time. The scale (j) increases with the inverse of the frequency: the weaker j, the faster the local oscillation. The darker the box, the more power is associated with it.

The wavelet coefficients images show the different points localized according to different frequency scales.

Maximum entropy is the most likely solution to the opposite problem of EEG (Rice., 1990). Therefore, the use of such a spatial model in the MEM framework is the key idea allowing MEM to be sensitive to the spatial extent of sources along the cortical surface (Grova et al., 2006). In addition, the wMEM method is sensitive to the spatial extent of the generators, even more so when the local spatial regularity is imposed in the plots, ie the standard MEM approaches in the time domain on the EEG data (Grova et al., 2006). Moreover, the standard MEM method is able to recover the spatial extent of sources more precisely and reliably than the other standard solvers (MNE, LORETA) (Lina et al., 2011). The wMEM methodology is a promising methodology that can be used to study all oscillatory models under healthy (C. Tallon et al., 1999; Grimault et al., 2009) or pathological conditions

## Conclusion:

Combined TMS-EEG is becoming an important tool for assessing cortical circuits and brain networks. FMRI studies performed during TMS may be able to identify zones of the brain activated by TMS and regions functionally connected to the stimulated sites. It will be also of a specific utility in the cases of the locations of the zones.

In the present study, the localization of the sources have been detailed which makes it possible to evaluate the localization of the brain generators by surface measurements and to determine the active zones in the different methods distributed following stimulation by TMS. Subsequently, three comparison metrics between the four inverse solutions were proposed.

Several methods have been proposed to solve the localization source problem in the human brain, such as dipolar methods, scanning methods, spatial filters and distributed methods. In the present study, the tests focused on three techniques of linear localization of distributed sources to solve the study problems which are MNE, dSPM and sLORETA and the nonlinear MEM method of the wMEM type based on the wavelet structure. Then, the impact of TMS technique on the EEG data was studied for four different algorithms: minimum estimate of the standard (MNE), dynamic statistical parametric mapping dSPM), standardized low resolution electromagnetic tomography(sLORETA) and the (wavelet based on the Maximum Entropy on the Mean (wMEM). Concerning the linear methods, the present work examined MNE, dSPM and sLORETA and for the nonlinear methods, the present study focused on the "wMEM" method. It is a probabilistic approach method based on the time-frequency domain. Such method renders an effective and efficient result in the case of the localization of points: presence of active zones whatever the frequency scale used (there are peaks located in all frequency scales). In addition, this method uses a scale specific covariance matrix and scale dependent shrinkage in the case of the presence of a data set.

In the present experiment, the comparison metrics show that sLORETA outperforms the wMEM, dSPM and MNE with the latter showing the least detection in activity mapping. As far as the notion of strength is concerned there is a presence of decrease in the strength of connectivity between the two hemispheres after using TMS in the majority of methods (MNE, dSPM, sLORETA and wMEM) of intra zones.Based on the present study result, on the one hand the TMS technique makes it possible to increase the number of active zones in the different brain regions (for example the percentage of detection of sLORETA goes from 1.71% before TMS to 5.15% after TMS), d " to increase inter-zone connectivity for all localization methods. On the other hand, it makes it possible to decrease intra-zone connectivity, and to increase the activation of zones at the level of the occipital lope. To conclude, four different solutions of distributed localization sources MNE, dSPM, sLORETA were compared to the nonlinear method wMEM using three comparison metrics: detection of active zones, strength of inter and intra zones connectivity, strength of cross-correlation to study the impact of TMS on EEG data.